%% file: NoisyGaussianLattice.tex
\documentclass[10pt,journal,twocolumn]{IEEEtran}
\usepackage{amsthm}
\usepackage{epsfig}
\usepackage{xfrac}
\usepackage{amsmath}
\usepackage{cases}
\usepackage{amssymb,mathrsfs,dsfont}
\usepackage{enumerate}
\usepackage{color}
\usepackage{textcomp}
\usepackage{tikz}
\usepackage{pgfplots}
\usepackage{mathtools}
\usepackage{scrextend}
\newtheorem{theorem}{Theorem}
\newtheorem{lemma}{Lemma}

\newtheorem{remark}{Remark}

\def\md{\mathbb}

\def\eps{\varepsilon}
\def\tn{\textnormal}
\def\wt{\widetilde}
\def\wh{\widehat}

\def\Expt{\md{E}}

\def\db{\mathrm{dB}}
\def\latticeN{N}
\def\SKn{k}
\def\SKN{K}
\def\V0{\mathcal{V}_0}
\def\Zeq{\bM{Z}_{\textrm{eq}}}
\def\dotleq{\stackrel{.}{\leq}}
\def\dotgeq{\stackrel{.}{\geq}}
\def\schemeind{i}

\newcommand{\bM}[1]{\boldsymbol{#1}}

\newcommand{\dfn}{ \stackrel{\tn{def}}{=} }

\newcommand{\moduloLattice}[1] {\mathbb{M}_{\Lambda}\left[#1\right]}
\newcommand{\quantLattice}[1] {\mathbb{Q}_{\Lambda}\left[#1\right]}

\newcommand{\snr}{\mathrm{SNR}}
\newcommand{\bsnr}{\mathrm{S}\wt{\mathrm{N}}\mathrm{R}}
\newcommand{\dsnr}{\Delta\snr}
\newcommand{\Pe}{p_e}
\newcommand{\Pmod}{p_\textrm{mod}}
\newcommand{\Pdec}{p_\textrm{dec}}
\newcommand{\latticeLoose}{L}

\usepackage[noadjust]{cite}

\begin{document}

\title{The Gaussian Channel with Noisy Feedback: Improving Reliability via Interaction}

\author{Assaf~Ben-Yishai and Ofer~Shayevitz
\thanks{The authors are with the Department of EE--Systems, Tel Aviv University, Tel Aviv, Israel \{assafbster@gmail.com, ofersha@eng.tau.ac.il\}. The research leading to these results has received funding from the European Research Council under the European Community's Seventh Framework Programme (FP7/2007-2013) / ERC grant agreement no. 639573, and from the Israel Science Foundation under grant agreement no. 1367/14.}
}

\maketitle

\begin{abstract}
Consider a pair of terminals connected by two independent (feedforward and feedback) \textit{Additive White Gaussian Noise} (AWGN) channels, and limited by individual power constraints. The first terminal would like to reliably send information to the second terminal at a given rate. While the reliability in the cases of no feedback and of noiseless feedback is well studied, not much is known about the case of noisy feedback. In this work, we present an interactive scheme that significantly improves the reliability relative to the no-feedback setting, whenever the feedback Signal to Noise Ratio ($\snr$) is sufficiently larger than the feedforward $\snr$. The scheme combines Schalkwijk-Kailath (S-K) coding and modulo--lattice analog transmission. 
\end{abstract}

\section{Introduction}\label{sec:introduction}

Feedback cannot improve the capacity of point-to-point memoryless channels \cite{ShannonFeedback}. Nevertheless, noiseless feedback can significantly simplify the transmission schemes and improve the error probability performance, see e.g. \cite{S-K_partII,horstein,PM_Transactions}. These elegant schemes fail however in the presence of arbitrarily small feedback noise, rendering them grossly impractical. This fact has been initially obseved in \cite{S-K_partII} for the AWGN channel, and further strengthened in \cite{KimNoisyAWGNFeedbackAllertor}. 

In a previous work \cite{SimpleInteractionAllerton2014} we presented a variation of the noiseless-feedback AWGN S-K scheme \cite{S-K_partII}, extending it to the case of noisy feedback. The scheme was based on the following observation: In each round, the receiver has some estimate of the message, and the transmitter needs to learn the associated estimation error in order to proceed. This estimation error can be conveyed in a power-efficient manner by using the knowledge of the message at the transmitter as side-information. The main focus of \cite{SimpleInteractionAllerton2014} was on the simplicity of the scheme in a fixed error probability regime, and side information was used by applying scalar modulo operations. This resulted in a major improvement of the capacity-gap in a relatively small number of rounds. 

The focus of this work is on the virtues of noisy feedback for increasing reliability. To that end, an asymptotic generalization of the scheme in \cite{SimpleInteractionAllerton2014} is introduced, applying the S-K scheme over blocks and replacing the scalar modulo with  multi-dimensional lattice modulo, as well as replacing Pulse Amplitude Modulation (PAM) used in \cite{SimpleInteractionAllerton2014} with a block code. An asymptotic error analysis is provided, using the Poltyrev exponent to account for modulo aliasing errors, and channel coding error exponents to account for the error of the block code. The resulting error exponent is computed and shown to surpass the sphere-packing bound of the feedforward channel for a wide range of rates and $\snr$ settings.

In \cite{KimNoisyAWGNFeedbackAllertor,BurnashevNoisyAWGNISIT}, the authors analyzed the reliability function of the AWGN  at zero rate for noisy \textit{passive} feedback, i.e. where the channel outputs are fed back without any processing. In \cite{KimActiveFB}, which is closer to our interactive setting, the reliability function of the AWGN at zero rate (two messages) with noisy \textit{active}  feedback has been considered. Specifically, it was shown that active feedback roughly quadruples the error exponent relative to passive feedback. The achievability result of \cite{KimActiveFB} is better than ours at zero rate. 

\section{Preliminaries}

We write $\log$ for base $2$ logarithm, and $\ln$ for the natural logarithm. We use the vector notation $x^n\dfn [x_1,\ldots,x_n]$ and boldface letters such as $\bM{x}$ to indicate vectors of size $\latticeN$. We write $a_n\dotgeq b_n$ to mean  $\liminf_{n\rightarrow\infty}\frac{1}{n}\ln\left(\frac{a_n}{b_n}\right)\geq 0$, and similarly define $\dotleq$ and $\doteq$.

\subsection{Lattice Properties}
\label{sec:lattice-properties}
\begin{enumerate}[(i)]
\item We denote a lattice of dimension  $\latticeN$ by $\Lambda=G\cdot\mathbb{Z}^\latticeN$ where $G$ is the generating matrix. 
\item $V(\Lambda)=|\det(G)|$ is the lattice cell volume.
\item We denote the nearest neighbor quantization of $\bM{x}$ to the lattice $\Lambda$ by $\quantLattice{\bM{x}}$.
\item We denote the fundamental Voronoi cell $\V0=\{\bM{x}:\quantLattice{\bM{x}}=\bM{0} \}$.
\item Modulo $\Lambda$ is defined as $\moduloLattice{\bM{x}}\dfn \bM{x}-\quantLattice{\bM{x
}}$.
\item $\moduloLattice{\cdot}$ satisfies the \textit{distributive law} : $\moduloLattice{\moduloLattice{\bM{x}}+\bM{y}}=\moduloLattice{\bM{x}+\bM{y}}$.
\item The volume to noise ratio (VNR) of a lattice in the presence of AWGN with variance $\sigma^2$ is $\mu\dfn V^{2/\latticeN}(\Lambda)/\sigma^2$. 
\item The normalized second moment of a lattice $\Lambda$ is  $G(\Lambda)\dfn \sigma^2(\Lambda)/V^{2/\latticeN}(\Lambda)$, where $\sigma^2(\Lambda)=\frac{1}{\latticeN}\Expt{(\|\bM{U} \|^2)}$ and $\bM{U}$ is uniformly distributed on $\V0$.
\end{enumerate}

\subsection{Joint Source Channel Coding (JSCC)}
\label{sec:joint-source-channel}
It is well known \cite{Goblick} that when a Gaussian source is conveyed over AWGN channel and quadratic distortion measure, analog transmission obtains the optimal distortion with minimal delay. The transmitter merely has to scale the source $Q$ to the channel input power constraint, and the receiver merely has to multiply by the appropriate Wiener coefficient in order to obtain the optimal linear estimate. This solution is a simple case of joint source channel coding (JSCC).

If side information related to the source $Q$ is present at the receiver the problem is known as the Wyner-Ziv problem \cite{WynerZiv}. Kochman and Zamir \cite{KochmanZamirJointWZWDP} gave the solution of JSCC with side information (related to the source and the channel) over an AWGN with quadratic distortion measure. They used analog transmission as \cite{Goblick} in conjunction with dithered modulo lattice operations that take care of the side information. 

Let us quickly quote survey their result in the case of side information only at the source. The source vector to be conveyed is $\bM{Q}+\bM{J}$ where the destination has $\bM{J}$ as side information. The channel is AWGN with input $\bM{X}$ noise $\bM{Z}$ and output $\bM{Y}$, i.e. $\bM{Y}=\bM{X}+\bM{Z}$. The transmitter sends: 
\begin{align*}
\bM{X} = \moduloLattice{\beta (\bM{J}+\bM{Q}) + \bM{V}}
\end{align*}
Where $\bM{V}$ is the dither vector uniformly distributed on $\V0$ the basic Voronoi cell of $\Lambda$ and commonly known at the transmitter and receiver. The receiver first calculates the temporary variable $\bM{T}$ as follows:
\begin{align*}
\bM{T} \dfn \alpha_C \bM{Y} - \bM{V} - \beta \bM{J} =\bM{X}+\Zeq-\bM{V}-\beta \bM{J}
\end{align*}
where the second transition is pedestrian by the definition of the equivalent noise $\Zeq$:
\begin{align*}
\Zeq \dfn -(1-\alpha_C)\bM{X}+\alpha_C\bM{Z} 
\end{align*}
The receiver now applies another modulo operation on $\bM{T}$ obtaining $\bM{U}$ as follows:
\begin{align*}
\bM{U} \dfn \moduloLattice{\bM{T}}
=\moduloLattice{\beta\bM{Q}+\Zeq}
\end{align*}
where the second transition is due to the distributive law on $\moduloLattice{\cdot}$. Now, if $\beta\bM{Q}+\Zeq \in \V0$, then  $\bM{U}=\beta\bM{Q}+\Zeq$. We show in the sequel that by appropriate parameter settings, the probability of the complementary event can be made exponentially small with respect to the lattice dimension $\latticeN$.

In  \cite{KochmanZamirJointWZWDP} a linear estimate of $\bM{Q}$: $\wh{\bM{Q}}=\frac{\alpha_S}{\beta} \bM{U}$ was obtained. However, in our scheme $\frac{\alpha_S}{\beta}$ naturally cancels out rendering the setting $\alpha_S$ immaterial. We note that setting $\alpha_C<1$ is common practice in many lattice problems, improving performance in lower $\snr$s and making $\Zeq$ non-Gaussian (which usually improves the error probability). For clarity of exposition we use in this work only $\alpha_C=1$.

\subsection{The  Schalkwijk-Kailath (S-K) Scheme}
\label{sec:schalkw-kail-sk}
The famous S-K scheme\cite{S-K_partII} for capacity achieving communication over AWGN can be interpreted using JSCC tools. The classic scheme encodes the message $W$ into a message point $\Theta$ using single dimensional modulation. At the end of the first step, $\SKn=1$, Terminal B sets its estimate of $\Theta$ to be  $\wh{\Theta}_{1}=Y_1$. In consequent steps $\SKn$, Terminal B feeds back $\wh{\Theta}_{\SKn}$ to Terminal A. At step $\SKn+1$, Terminal A extracts the estimation error $\eps_{\SKn}=\wh{\Theta}_{\SKn}-\Theta$ and conveys it to Terminal B by JSCC. Namely, Terminal A sends $X_{\SKn}=\alpha_{\SKn+1}\eps_{\SKn}$ where $\alpha_{\SKn+1}$ is set so that to meet the channel input power constraint and Terminal B linearly estimates $\wh{\eps}_{\SKn}=\beta_{\SKn+1}Y_{\SKn+1}$ where $\beta_{\SKn+1}$ is set so that to minimize the \textit{Mean Squared Error} (MSE). Having $\wh{\eps}_{\SKn}$,
 Terminal B now advances its estimate by $\wh{\Theta}_{\SKn+1}=\wh{\Theta}_{\SKn+1}-\wh{\eps}_{\SKn}$. Finally, at step $\SKN$, Terminal B decodes $W$ from $\wh{\Theta}_{\SKN}$.

For the sake of analysis it is convenient to observe the series of channels from $\Theta$ to $\wh{\Theta}_{\SKn}$. These are Gaussian channels whose noise variance is $\sigma^2$ at $\SKn=1$, and it is easy to see that optimizing over $\alpha_{\SKn}$ and $\beta_{\SKn}$ reduces the noise variance by $1+\snr$ at every step. So, at the final step $\SKN$ we have a channel whose $\snr$ is $\snr(1+\snr)^{\SKN-1}$. At this step it can be shown that mapping $W$ into $\Theta$ using PAM and giving a Gaussian analysis of the error probability, can yield a rate arbitrarily close to the channel capacity by taking a sufficiently large $\SKN$.
 
\subsection{Error Exponents}
Consider the case where a lattice point $\bM{X}\in \Lambda$ is sent over an AWGN channel $\bM{Y}=\bM{X}+\bM{Z}$ and the decoder estimates $\wh{\bM{X}}(\bM{Y})$ according to an \textit{Maximum Likelihood} (ML) decoding rule. Then there exist lattices whose probability of decoding error is exponentially upper bounded by $\Pr(\wh{\bM{X}}(\bM{Y})\neq \bM{X}) \dotleq e^{-\latticeN E_p(\frac{\mu}{2\pi e})}$.
where $\mu$ is the VNR w.r.t the lattice and the channel noise variance and $E_p(\cdot)$ is the Poltirev error exponent given by \cite{RamiLattices,GoodLattices}:
\begin{align*}
%\label{eq:Ep}
E_p(x) = 
\begin{cases}
\frac{1}{2}\left(x-1-\ln (x)\right) & \text{if } 1<x\leq 2 \\
\frac{1}{2}\left(\ln(x)+\ln(\frac{e}{4}) \right) & \text{if } 2<x\leq 4 \\
\frac{1}{8}x & \text{if } x>4 \\
\end{cases}
\end{align*}

For channel coding over AWGN with $\snr$ and with rate $R$, there exist block codes of length $\latticeN$ whose average error probability (averaged over the messages) under ML decoding is exponentially upper bounded by $\Pr(\wh{\bM{X}}(\bM{Y})\neq \bM{X}) \dotleq e^{-\latticeN E_r(R)}$ where $E_r(\snr,R)$is given by \cite{GallagerIT} :
\begin{align*}
%\label{eq:Er}
E_r(\snr,R) = 
\begin{cases}
E_{sp}(\snr,R) & \text{if } R_{rc}<R\leq C  \\
E_{rc}(\snr,R) & \text{if } R_{ex}<R\leq R_{rc} \\
E_{ex}(\snr,R) & \text{if } 0 < R \leq R_{ex} \\
\end{cases}
\end{align*}
The boundaries between the regions are as follows. The \textit{Shannon capacity} is $C\dfn\frac{1}{2}\log(1+\snr)$. The \textit{critical rate} is $R_{cr}\dfn \sfrac{1}{2}\log\left(\sfrac{1}{2}+\sfrac{\snr}{4}+\sfrac{1}{2}\sqrt{1+\sfrac{\snr^2}{4}}\right)$. The \textit{expurgation rate} is $R_{ex}\dfn \sfrac{1}{2}\log\left(\sfrac{1}{2}+\sfrac{1}{2}\sqrt{1+\sfrac{\snr^2}{4}}\right)$. 

The error exponent in the \textit{sphere packing} region is:
\begin{align*}
%\label{eq:Esp}
E_{sp}(\snr,R)&=\frac{\snr}{4\beta}\left(\beta+1-(\beta-1)\sqrt{1+\frac{4\beta}{\snr(\beta-1)}}\right)\nonumber \\
&+\frac{1}{2}\ln\left(\beta- \frac{\snr(\beta-1)}{2}\sqrt{1+\frac{4\beta}{\snr(\beta-1)}}\right)
\end{align*}
where $\beta=2^{2R}$. In the \textit{random coding} region:
\begin{align*}
E_{rc}(\snr,R)&=1-\beta+\frac{\snr}{2}+\frac{1}{2}\log\left(\beta-\frac{\snr}{2}\right)\\
&-\frac{1}{2}\log(\beta)-\log(2)R
\end{align*}
where now $\beta=2e^{2R_{cr}}$. Lastly, in the expurgation region:
\begin{align*}
E_{ex}(\snr,R)&=\frac{\snr}{4}\left[1-\sqrt{1-2^{-2R}} \right]
\end{align*}

 It is also possible to show, using some pedestrian algebra, that for all rates $0<R<C$, $E_{sp}(\snr,R)$ coincides with the asymptotic expression of Shannon's sphere packing bound for AWGN \cite{ShannonSpherePacking59}. Hence, it is also an upper bound for the reliability function, and thus the bound is tight above the critical rate. 

\section{Setup}
Our setup is defined as follows. The feedforward and feedback channels connecting Terminal A to Terminal B and vice versa respectively, are AWGN channels given by 
\begin{align*}
Y_n=X_n+Z_n,\quad \wt{Y}_n=\wt{X}_n+\wt{Z}_n.
\end{align*}
 Where $X_n, Y_n$ (resp. $\wt{X}_n,\wt{Y}_n$) are the input and output of the feedforward (resp. feedback) channel at time $n$ respectively. The feedforward (resp. feedback) channel noise $Z_n\sim \mathcal{N}(0,\sigma^2)$ (resp. $\wt{Z}_n\sim \mathcal{N}(0,\wt{\sigma}^2)$) is independent of the input $X_n$ (resp. $\wt{X}_n$), and constitutes an i.i.d. sequence. The feedforward and feedback noise processes are mutually independent. 

Terminal A is in possession of a message $W\sim \textrm{Uniform}([M])$, to be described to Terminal B over $N$ rounds of communication. To that end, the terminals can employ an interactive scheme defined by a pair of functions $(\varphi,\wt{\varphi})$ as follows: At time $n$, Terminal A sends a function of its message $W$ and possibly of past feedback channel outputs over the feedforward channel, i.e., 
\begin{align*}
  X_n=\varphi(W,\wt{Y}^{n-1}).
\end{align*}
Similarly, Terminal B sends function of its past observations to Terminal A over the feedback channel, i.e., 
\begin{align*}
  \wt{X}_n=\wt{\varphi}(Y^n).
\end{align*}
\begin{remark}
The dependence of $\varphi$ and $\wt{\varphi}$ on $n$ is suppressed. In general, we allow these functions to further depend on common randomness shared by the terminals.
\end{remark}

We assume that Terminal A (resp. Terminal B) is subject to a power constraint $P$ (resp. $\wt{P}$), namely
\begin{align*}
\sum_{n=1}^N\mathbb{E}(X_n^2) \leq N\cdot P, \quad \sum_{n=1}^N\mathbb{E}(\wt{X}_n^2) \leq N\cdot \wt{P} .
\end{align*}
We denote the feedforward (resp. feedback) $\snr$ by $\snr\dfn\frac{P}{\sigma^2}$ 
(resp.  $\bsnr\dfn \frac{\wt{P}}{\wt{\sigma}^2}$). The ratio between the feedback $\snr$ and the feedforward $\snr$ is denoted by $\dsnr\dfn\frac{\bsnr}{\snr}$. We implicitly assume that $\dsnr>1$.

An interactive scheme $(\varphi,\wt{\varphi})$ is associated with a rate $R\dfn \frac{\log{M}}{N}$ (in bits) and an error probability $\Pe(N,R)$, which is the probability that Terminal B errs in decoding the message $W$ at time $N$, under the optimal decision rule. We say that an error exponent $E(R)$ is achievable if there exists a sequence of interactive coding schemes indexed by $N$ with rate at least $R$, such that $p_e(N,R) \dotleq e^{-NE(R)}$.

\section{Description of the scheme}
\label{sec:description-scheme}
In Subsection \ref{sec:schalkw-kail-sk} we discussed the S-K scheme and described its feedforward transmission as a JSCC of the estimation error. It was assumed that Terminal A knows the estimation error, which is made possible by Terminal B sending back its estimate $\wh{\Theta}_{\SKn}$, and Terminal A in turn subtracting $\Theta$ to obtain  $\eps_{\SKn}=\wh{\Theta}_{\SKn}-\Theta$. This procedure holds if the feedback is noiseless and fails if the feedback is noisy. In the latter case, it was observed in \cite{SimpleInteractionAllerton2014} that the transmission from Terminal B to Terminal A can be regarded as JSCC with side information. Namely, Terminal B wished to convey $\eps_{\SKn}$ to Terminal A, but knowing only  $\wh{\Theta}_{\SKn} =\Theta+\eps_{\SKn}$ whereas Terminal A knows $\Theta$ and can use it as side information. The scheme described in \cite{SimpleInteractionAllerton2014} used scalar modulo operations and took advantage of the noisy feedback in order to reduce the capacity gap, maintaining the simplicity of the original S-K scheme.

The use of scalar modulo operation benefits from simplicity and low delay, at the price of modulo error which is bounded away from zero. As shown in Subsection \ref{sec:joint-source-channel}, the error probability can be made to approach zero using modulo-lattice operations in the limit of large dimension. This provides motivation to the following modifications of our scalar scheme: 
\begin{enumerate}
\item Replace the scalar interval lattice with a lattice $\Lambda$ of dimension $\latticeN$. 
\item Replace the scalar PAM mapping of the message point $W\rightarrow \Theta$ with an AWGN block code of the same dimension  $\latticeN$,namely $W\rightarrow \bM{\Theta}$.
\item Use block code and lattice error exponents for the analysis of the aggregate error probability incurred by the  associated high-dimensional extension of our scalar scheme, where interaction takes place on a block-wise basis. 
\end{enumerate}

It should be noted that the feedback operations (i.e. the modulo-lattice operations) requires the knowledge of an entire vector of length $\latticeN$, and cannot be implemented on the fly. Moreover, the modulo-lattice result  requires $\latticeN$ channel uses to be transmitted. To accommodate this inherent delay we use two interlaced block-wise schemes. Having two schemes each using $\SKN$ rounds requires $2\SKN$ blocks of length $\latticeN$. For simplicity, we use double indexing for the blocks. The block index $l$ is represented by a pair of indices ($\SKn$,$j$) so that $l=2(\SKn-1)+j$. More explicitly, this notation defines
\begin{align*}
\bM{X}_{\SKn}^{j} \dfn \left[{X}_{(2(\SKn-1)+j-1)\latticeN+1},\ldots,X_{(2(\SKn-1)+j-1)\latticeN+\latticeN}\right]
\end{align*}
We denote the round index by $\SKn\in [\SKN]$ and the scheme index by $\schemeind\in\{1,2\}$. The feedforward of round $\SKn$ and scheme $\schemeind$ is sent over the block pertaining to indices $(\SKn,\schemeind)$, and the corresponding feedback is sent over the block pertaining to indices $({\SKn},\schemeind+1)$.

Let us now give a description of scheme for ${\schemeind}\in\{1,2\}$. The setting of the parameters $\alpha,\beta_{\SKn},\gamma_{\SKn}$ will be discussed in the sequel. The dither variables $\bM{V}_{\SKn}^{\schemeind}$ are i.i.d. and uniformly distributed on $\V0$.
\begin{enumerate}[(A)]
\item Initialization:
\begin{enumerate}[]
\item \textbf{Terminal A:} Map the message $W^{\schemeind}$ to codeword $\bM{\Theta}^{\schemeind}$ using a codebook for AWGN with average power $P$. 
\item \textbf{Terminal A $\Rightarrow$ Terminal B:} 
  \begin{itemize}
  \item Send $\bM{X}_1^{\schemeind}=\bM{\Theta}^{\schemeind}$
    \item Receive $\bM{Y}_1^{\schemeind}=\bM{X}_1^{\schemeind}+\bM{Z}_1^{\schemeind}$
  \end{itemize}  

\item \textbf{Terminal B:} Initialize the $\bM{\Theta}^i$ estimate
to $\wh{\bM{\Theta}}_1^{\schemeind}=\bM{Y}_1^{\schemeind}$. 
\end{enumerate}
\item Iteration:
\begin{enumerate}[]
\item \textbf{Terminal B $\Rightarrow$ Terminal A:} 
  \begin{itemize}
  \item Given the $\bM{\Theta}^{\schemeind}$ estimate $\wh{\bM{\Theta}}^{\schemeind}_{\SKn}$, compute and send in the following block
    \begin{align*}
      \wt{\bM{X}}_{\SKn}^{\schemeind+1}= \moduloLattice{\gamma_n\wh{\bM{\Theta}}^{\schemeind}_{\SKn}
+\bM{V}_{\SKn}^{\schemeind}}
    \end{align*}
  \item Receive $\wt{\bM{Y}}_{\SKn}^{\schemeind+1}=\wt{\bM{X}}_{\SKn}^{\schemeind+1} + \wt{\bM{Z}}_{\SKn}^{\schemeind+1}$
  \end{itemize}  
\item \textbf{Terminal A:} Extract a noisy scaled version of
estimation error $\bM{\eps}_{\SKn}^{\schemeind}$:
\begin{align*}%\label{eq:noisy_estimate}
\wt{\bM{\eps}}_{\SKn}^{\schemeind}=\moduloLattice{\wt{\bM{Y}}_{\SKn}^{\schemeind+1}-\gamma_n{\bM{\Theta}}^{\schemeind}-\bM{V}_{\SKn}^{\schemeind}} 
\end{align*}
Note that $\wt{\bM{\eps}}_{\SKn}^{\schemeind}  = \gamma_n \bM{\eps}_{\SKn}^{\schemeind} + \wt{\bM{Z}}_{\SKn}^{\schemeind+1}$, unless a \textit{modulo-aliasing error} occurs. 
 \item \textbf{Terminal A $\Rightarrow$ Terminal B:} 
  \begin{itemize}
  \item Send a scaled version of $\wt{\bM{\eps}}_{\SKn}^{\schemeind}$:
$\bM{X}_{\SKn+1}^{\schemeind}=\alpha \wt{\bM{\eps}}_{\SKn}^{\schemeind}$, where $\alpha$ is set so that to meet the input power constraint $P$ (computed later). 
  \item Receive $\bM{Y}_{\SKn+1}^{\schemeind}=\bM{X}_{\SKn+1}^{\schemeind}+\bM{Z}_{\SKn+1}^{\schemeind}$
  \end{itemize}  
\item \textbf{Terminal B:} 
Update the $\bM{\Theta}^{\schemeind}$ estimate
 $\wh{\bM{\Theta}}_{\SKn+1}^{\schemeind}=\wh{\bM{\Theta}}_{\SKn}^{\schemeind}-\wh{\bM{\eps}}_{\SKn}^{\schemeind}$, where  
\begin{align}
\label{eq:etahat}
\wh{\bM{\eps}}_{\SKn}^{\schemeind}=\beta_{\SKn+1}\bM{Y}_{\SKn+1}^{\schemeind}  
\end{align}
is the MMSE estimate of $\bM{\eps}_\SKn$. The optimal selection of $\beta_{\SKn}$ is described in the sequel. 
\end{enumerate}
\item Decoding: 
After the reception of block $\bM{Y}_{\SKN}^{\schemeind}$ the receiver decodes the message $\wh{W}^{\schemeind}\left(\bM{Y}_{\SKN}^{\schemeind}\right)$ using an ML decision rule w.r.t. the codebook. 
\end{enumerate}

\section{Error analysis and Parameter Setting}
As elaborated above, decoding of the two interlaced schemes produces $\wh{W}^{\schemeind}\left(\bM{Y}_{\SKN}^{\schemeind}\right)$ and an error occurs if either of the decoded messages is not equal to its corresponding sent message $W^\schemeind$. It is important to note that due to the modulo-lattice operations in the feedback, the additive noise corrupting $\bM{Y}_{\SKN}^{\schemeind}$ is not Gaussian. However, a Gaussian analysis can be used to bound the error probability as we show herein.

For any $\SKn\in\{1,\ldots, \SKN-1\}$ we define $E_{\SKn}^{\schemeind}$ as the event the feedback decoding results in modulo-aliasing error, i.e.
\begin{align*}
%\label{eq:moderror}
E_{\SKn}^{\schemeind}\dfn\left\{ \gamma_n \bM{\eps}_{\SKn}^{\schemeind} + \wt{\bM{Z}}_{\SKn}^{\schemeind+1}\notin \V0\right\}.
\end{align*}
We define $E_{\SKN}^{\schemeind}$ as the decoding error at the final decoding step
\begin{align*}
E_{\SKN}^{\schemeind}=\left\{\wh{W}^\schemeind(\bM{Y}_{\SKN}^{\schemeind}) \neq W^\schemeind\right\}
\end{align*}

In order to use the Gaussian analysis we introduce the following upper bound for the error probability: 
\begin{align}\label{eq:err_union}
\Pe^\schemeind \leq \Pr\left(\bigcup_{\SKn=1}^\SKN E_{\SKn}^{\schemeind}\right).
\end{align}
The inequality stems from the fact that a modulo-aliasing error does not necessarily cause a decoding error.

To proceed, we define the \textit{coupled system} as a system that is fed by the same message and experiences the (sample-path) exact same noises, with the only difference being that no modulo operations are implemented at neither of the terminals. Clearly, the coupled system violates the power constraint at Terminal B. However, given the message $W^{\schemeind}$, all the random variables in the coupled system are jointly Gaussian, and in particular, the estimation errors $ \bM{\eps}_{\SKn}^{\schemeind}$ in that system are Gaussian for $\SKn=1,\ldots,\SKN$. Moreover, it is easy to see that the estimation errors are \textit{sample-path identical} between the original system and the coupled system until the first modulo-aliasing error occurs. To be precise we quote \cite[Lemma 1]{SimpleInteractionAllerton2014}:
\begin{lemma}
Let  $\wt{\Pr}$ denote the probability operator in for the coupled process. Then for any $\SKN>1$:
\begin{align*}
\Pr\left(\bigcup_{\SKn=1}^\SKN E_\SKn^{\schemeind} \right) =  \wt{\Pr}\left(\bigcup_{\SKn=1}^\SKN E_\SKn^{\schemeind} \right).
\end{align*}
 \end{lemma}

Combining the above with \eqref{eq:err_union} and applying the union bound in the coupled system, we obtain
\begin{align*}
\Pe\leq\sum_{\schemeind=1}^{2}\sum_{\SKn=1}^{\SKN}\wt{\Pr}\left( E_\SKn^\schemeind\right).
\end{align*}
Calculating the above probabilities now involves only Gaussian random variable, which significantly simplifies the analysis.

From this step on, we perform an asymptotic exponential analysis. We note that the sums of probabilities are exponentially dominated by the maximal summand, therefore both interlaced schemes $\schemeind\in\{1,2\}$ are set to be identical, and set the parameters such that all modulo-aliasing error probabilities are the same. Hence
\begin{align*}
\Pe\leq 2\left[(K-1)\wt{\Pr}\left( E_1^\schemeind\right)+\wt{\Pr}\left( E_\SKN^\schemeind\right)\right]
\doteq \wt{\Pr}\left( E_1^1\right)+\wt{\Pr}\left( E_\SKN^1\right).
\end{align*}
Defining $\Pmod\dfn\wt{\Pr}\left( E_1^1\right)$ and  $\Pdec\dfn\wt{\Pr}\left( E_\SKN^1\right)$ yields 
\begin{align*}
\Pe\dotleq \Pmod+\Pdec
\end{align*}
We now set the lattice second moment to equal the feedback power constraint $\sigma^2(\Lambda)=\wt{P}$ (and guarantee  that this is the feedback transmission power by dithering). The modulo-aliasing error event is the event where
\begin{align}
\label{eq:moduloError}
\gamma_n \bM{\eps}_{\SKn}^{\schemeind} + \wt{\bM{Z}}_{\SKn}^{\schemeind+1}\notin \V0
\end{align}
By the coupling argument, we can assume for our bounding analysis that the LHS above is Gaussian. The \textit{looseness} $\latticeLoose$ of the lattice is defined by the power ratio of the RHS and LHS of \eqref{eq:moduloError}, i.e., 
\begin{align*}
\latticeLoose \dfn \frac{\wt{P}}{\gamma_\SKn^2\sigma_\SKn^2+\wt{\sigma}^2}.
\end{align*} 
By the definitions in Subsection~\ref{sec:lattice-properties}: $\latticeLoose = \mu(\Lambda)\cdot G(\Lambda)$. By \cite[Theorem 5]{GoodLattices} there exist lattices that asymptotically attain both $G(\Lambda)= \frac{1}{2\pi e} + o(1)$ and the Poltyrev exponent, so we can set $\mu=2 \pi e\latticeLoose + o(1)$, then $E_p(\frac{\mu}{2\pi e})=E_p(\latticeLoose)$ and 
\begin{align*}
\Pmod \dotleq e^{-N E_p(\latticeLoose)}.
\end{align*}

In the next step we send $\bM{X}_{\SKn+1}^{\schemeind}=\alpha \wt{\bM{\eps}}_{\SKn}^{\schemeind}$, where $\alpha$ is set so that to meet the input power constraint $P$, i.e. $\alpha = \sqrt{\latticeLoose\frac{P}{\wt{P}}}$. The channel output in the next round is thus 
\begin{align*}
\bM{Y}_{\SKn+1}^{\schemeind}=
\alpha\gamma_n \bM{\eps}_{\SKn}^{\schemeind} +\alpha \wt{\bM{Z}}_{\SKn}^{\schemeind+1}
+\bM{Z}_{\SKn+1}^{\schemeind}.
\end{align*}
Setting $\beta_{\SKn+1}$ in  \eqref{eq:etahat} to the optimal Wiener coefficient, one can easily calculate the evolution of the estimation variance $\sigma_{\SKn}^2\dfn\frac{1}{N}\Expt||\bM{\eps}_\SKn^{\schemeind} ||^2$. We now observe that the channel from $\bM{\Theta}^{\schemeind}$ to $\bM{\Theta}^{\schemeind}+\bM{\eps}_\SKn^{\schemeind}$ is in fact a vector of independent parallel  AWGN channels each with a noise variance $\sigma_{\SKn}^2$. Namely, after $\SKN$ rounds, we have $\latticeN$ instances of independent AWGN channels each with $\snr$ given by 
\begin{align*}
%label{eq SNRK}
\snr_\SKN(\latticeLoose)\dfn \snr\cdot\left(1+\snr\frac{1-\latticeLoose\bsnr^{-1}}{1+\latticeLoose\dsnr^{-1}}\right)^{\SKN-1}.
\end{align*}
Therefore, we can now map the message  $W^{\schemeind}$ into $\bM{\Theta}^{\schemeind}$ using a Gaussian codebook of block length $N$ and rate $K\cdot R$ to obtain 
\begin{align*}
\Pdec\dotleq e^{-\latticeN  E_r\left(\snr_\SKN(\latticeLoose),\SKN\cdot R\right)}.
\end{align*}
Note that the rate $K\cdot R$ is chosen such that the overall rate (over $K$ rounds) is $R$. We therefore immediately obtain the following. 

\begin{theorem}
The error probability attained by our suggested interactive scheme is upper bounded by $\Pe \dotleq e^{-\latticeN E_{FB}(R)}$, where
\begin{align}
\label{eq:EFB}
E_\textrm{FB}(R)&\dfn\max_{\SKN\in\mathbb{N},L\geq 1} \left\{\tfrac{\min \left\{ E_r ( \snr_{\SKN}(\latticeLoose),\SKN R ),E_p(\latticeLoose) \right\}}{2\SKN}\right\}
\end{align}
\end{theorem}

Note that the division by $2\SKN$ in due to normalization of the error exponents by the actual code length which is $2\latticeN\SKN$. The trade-off is now clear: setting the lattice looseness $\latticeLoose$ to be large reduces $\Pmod$ but also reduces $\snr_{\SKN}(L)$ hence enlarging $\Pdec$, and vice versa. Due to the monotonicity of $E_r (\snr_{\SKN}(\latticeLoose),\SKN\cdot R ),E_p(\latticeLoose)$ in $\latticeLoose$, a numerical solution to \eqref{eq:EFB} can be easily found.

\section{Discussion}
\label{results}
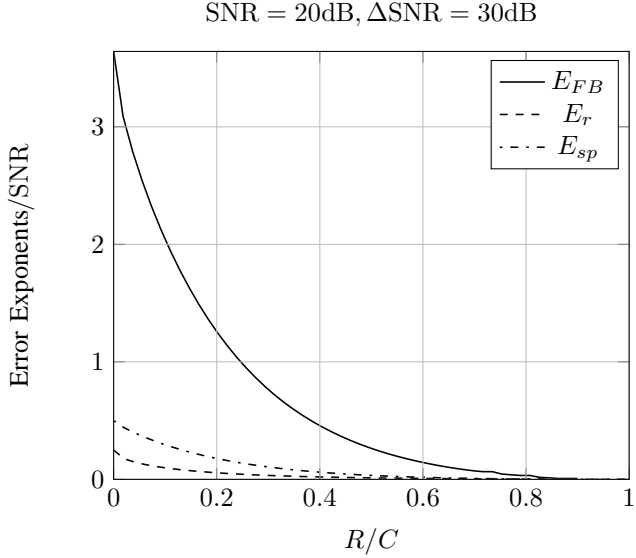
\begin{figure}
\input{EFB_snr20_dsnr30.tex}
\caption{\label{fig:EFB_snr20_dsnr30}
Error exponents with and without feedback for $\snr=20\db$ and $\dsnr=30\db$}
\end{figure}
Numerical evaluation of  $E_{sp}$,  $E_{r}$ and $E_{FB}$ for $\snr=20\db$ and $\dsnr = 30\db$ is depicted in Fig. \ref{fig:EFB_snr20_dsnr30}. It is clear that in this scenario our scheme improves the error exponents for most rates below capacity. 

It is now constructive to give an approximation for high $\snr$, namely $\snr\gg 1$. It is easy to see that for $\snr\gg 1$:
\begin{align*}
\snr_\SKN(\latticeLoose) \geq \frac{\left(\bsnr-\latticeLoose\right)^\SKN}{\dsnr\cdot\latticeLoose^{\SKN-1}}\left(1 + o(1)\right).
\end{align*}
We would now like to set $E_r ( \snr_{\SKN}(\latticeLoose),R\SKN))=E_p(\latticeLoose)$ and solve for $\latticeLoose$. Assuming both $E_r$ and $E_p$ are in their expurgation regions (as shall be verified later) we would like to solve:
\begin{align*}
%\label{eq:Leq}
\frac{1}{4}\frac{\left(\bsnr-\latticeLoose\right)^\SKN}{\dsnr\cdot\latticeLoose^{\SKN-1}}
\eta(R\SKN) =\frac{1}{8}\latticeLoose.
\end{align*}
where $\eta(R\SKN)\dfn 1-\sqrt{1-2^{-R\SKN}}$. The solution yields 
$L^*=\bsnr/(1+(\tfrac{1}{2}\eta(R\SKN)\dsnr)^{{\frac{1}{K}}})$. Plugging it in \eqref{eq:EFB} yields for any 
$\SKN>1$:
\begin{align}
\label{eq:EFBbound}
E_{FB}(R)\geq \frac{\snr\cdot\dsnr\cdot\eta(R\SKN) }{16 K \left(1+(\tfrac{1}{2}\eta(R\SKN)\dsnr)^{{\frac{1}{K}}}\right)}\left(1 + o(1)\right)
\end{align}
At $R=0$ an optimization on $K$ is possible, yielding $K^*$:
\begin{align*}
K^*= 0.78\cdot\ln\left(\tfrac{1}{2}\dsnr\right) \approx 0.18\cdot\dsnr_\db - 0.54.
\end{align*}
So either (best of) $\lceil K^* \rceil$ and $\lfloor K^* \rfloor$ can be plugged in \eqref{eq:EFBbound} giving a bound for $E_{FB}$. This bound holds as long as this $K$ and $L^*$ both satisfy the expurgation region assumptions: $L>4$ and $\SKN R<R_{cr}( \snr_{\SKN}(\latticeLoose))$.

For rates outside this region one can simply use $\frac{1}{2K}E_r(\snr_\SKN(\latticeLoose),\SKN R)$ with $\latticeLoose$ and $K$ found at the highest rate in the expurgation region.
\section{Acknowledgment}
The authors would like to thank Uri Erez for his help, and Ram Zamir for giving the initial motivation that lead to this work. 

\bibliographystyle{IEEEbib}
\bibliography{bibtex_references}

\end{document}

%% file: EFB_snr20_dsnr30.tex
% This file was created by matplotlib v0.1.0.
% Copyright (c) 2010--2014, Nico Schlömer <nico.schloemer@gmail.com>
% All rights reserved.
% 
% The lastest updates can be retrieved from
% 
% https://github.com/nschloe/matplotlib2tikz
% 
% where you can also submit bug reports and leavecomments.
% 
\begin{tikzpicture}

\begin{axis}[
title={$\mathrm{SNR}=20\mathrm{dB}, \Delta\mathrm{SNR}=30\mathrm{dB}$},
xlabel={$R/C$},
ylabel={$\textrm{Error Exponents}/\mathrm{SNR}$},
xmin=0, xmax=1,
ymin=0, ymax=3.64161943162664,
axis on top,
xmajorgrids,
ymajorgrids,
legend entries={{$E_{FB}$},{$E_r$},{$E_{sp}$}}
]
\addplot [semithick, black]
coordinates {
(0,3.64161943162664)
(0.0183633691077923,3.08834999201021)
(0.0367267382155846,2.79240990390203)
(0.0550901073233769,2.54362461493436)
(0.0734534764311692,2.32355434647096)
(0.0918168455389615,2.12509438363935)
(0.110180214646754,1.94451987876424)
(0.128543583754546,1.77947900790189)
(0.146906952862338,1.62827876150929)
(0.165270321970131,1.48958270872176)
(0.183633691077923,1.36226920801331)
(0.201997060185715,1.24536051516106)
(0.220360429293508,1.13798471614021)
(0.2387237984013,1.03935425783482)
(0.257087167509092,0.94875207982597)
(0.275450536616884,0.865523055117472)
(0.293813905724677,0.789067074639644)
(0.312177274832469,0.718893246784472)
(0.330540643940261,0.655001132387468)
(0.348904013048054,0.596308312649693)
(0.367267382155846,0.542393050779029)
(0.385630751263638,0.492867976897181)
(0.40399412037143,0.447377456247076)
(0.422357489479223,0.405594697951345)
(0.440720858587015,0.367219689351314)
(0.459084227694807,0.331976974302206)
(0.4774475968026,0.299613741449767)
(0.495810965910392,0.269897834730231)
(0.514174335018184,0.242616367651103)
(0.532537704125977,0.217718210886723)
(0.550901073233769,0.195192282470605)
(0.569264442341561,0.174517068497911)
(0.587627811449353,0.155545030672754)
(0.605991180557146,0.138140965020283)
(0.624354549664938,0.122181889312794)
(0.64271791877273,0.107880429184214)
(0.661081287880523,0.0947961351279818)
(0.679444656988315,0.0833661531762053)
(0.697808026096107,0.0728317676659354)
(0.716171395203899,0.0654492892316608)
(0.734534764311692,0.0654492783804874)
(0.752898133419484,0.0452242487709593)
(0.771261502527276,0.0389175285055906)
(0.789624871635069,0.0331033912806814)
(0.807988240742861,0.0331033821527459)
(0.826351609850653,0.0171456418307359)
(0.844714978958445,0.0128631793637603)
(0.863078348066238,0.00741557096318513)
(0.88144171717403,0.00741557238729444)
(0.899805086281822,0.00741555184553375)

};
\addplot [semithick, black, dashed]
coordinates {
(0,0.25)
(0.0204081632653061,0.175047241690995)
(0.0408163265306122,0.146410153599596)
(0.0612244897959184,0.125966893643821)
(0.0816326530612245,0.10993111647683)
(0.102040816326531,0.0967887224985049)
(0.122448979591837,0.085739111342036)
(0.142857142857143,0.0762925845613281)
(0.163265306122449,0.0681211439362944)
(0.183673469387755,0.0609910292141385)
(0.204081632653061,0.0547279820861153)
(0.224489795918367,0.0491976234659353)
(0.244897959183673,0.0442935656650247)
(0.26530612244898,0.0399298047193856)
(0.285714285714286,0.0360356316432481)
(0.306122448979592,0.0325521020784093)
(0.326530612244898,0.02942951088534)
(0.346938775510204,0.0266255378499022)
(0.36734693877551,0.0241038552010617)
(0.387755102040816,0.0218330612718692)
(0.408163265306123,0.0197858497893703)
(0.428571428571429,0.0179383528543715)
(0.448979591836735,0.0162696142644458)
(0.469387755102041,0.0147611622318599)
(0.489795918367347,0.0133966589993963)
(0.510204081632653,0.0121616107326852)
(0.530612244897959,0.0110431252261706)
(0.551020408163265,0.0100297079511448)
(0.571428571428571,0.00911108915811973)
(0.591836734693878,0.00827807636192676)
(0.612244897959184,0.00752242774899906)
(0.63265306122449,0.00683674296436401)
(0.653061224489796,0.00621436843935291)
(0.673469387755102,0.00564931496548959)
(0.693877551020408,0.00513618564534515)
(0.714285714285714,0.00466180418344129)
(0.734693877551021,0.00419087351845748)
(0.755102040816327,0.00371994285347368)
(0.775510204081633,0.00324901218848988)
(0.795918367346939,0.00277808152350608)
(0.816326530612245,0.00230715085852228)
(0.836734693877551,0.00183622019353848)
(0.857142857142857,0.00137092083145563)
(0.877551020408163,0.000972824636318401)
(0.897959183673469,0.000652855425612568)
(0.918367346938776,0.000403988348403636)
(0.938775510204082,0.000219830393692436)
(0.959183673469388,9.45635256950274e-05)
(0.979591836734694,2.28929431850364e-05)
(1,5.55111512312578e-17)

};
\addplot [semithick, black, dash pattern=on 1pt off 3pt on 3pt off 3pt]
coordinates {
(0,0.5)
(0.0204081632653061,0.439044059054141)
(0.0408163265306122,0.395603651748763)
(0.0612244897959184,0.357126980048727)
(0.0816326530612245,0.322542340088493)
(0.102040816326531,0.291305936304351)
(0.122448979591837,0.263033430944089)
(0.142857142857143,0.237415928932065)
(0.163265306122449,0.214190711497736)
(0.183673469387755,0.193128073234132)
(0.204081632653061,0.174024195396246)
(0.224489795918367,0.156696697311621)
(0.244897959183673,0.140981542545627)
(0.26530612244898,0.126730709562539)
(0.285714285714286,0.11381033631081)
(0.306122448979592,0.102099183299575)
(0.326530612244898,0.0914873257213958)
(0.346938775510204,0.0818750196337951)
(0.36734693877551,0.0731717062885707)
(0.387755102040816,0.0652951298290451)
(0.408163265306123,0.0581705503929223)
(0.428571428571429,0.0517300390283018)
(0.448979591836735,0.0459118437546104)
(0.469387755102041,0.0406598181383061)
(0.489795918367347,0.0359229052287566)
(0.510204081632653,0.0316546708059946)
(0.530612244897959,0.0278128807479658)
(0.551020408163265,0.0243591180056565)
(0.571428571428571,0.0212584352289718)
(0.591836734693878,0.018479039547029)
(0.612244897959184,0.0159920063958968)
(0.63265306122449,0.0137710196203703)
(0.653061224489796,0.011792135365321)
(0.673469387755102,0.0100335675247389)
(0.693877551020408,0.00847549273898117)
(0.714285714285714,0.00709987312772041)
(0.734693877551021,0.00589029512137369)
(0.755102040816327,0.00483182291038511)
(0.775510204081633,0.00391086517204818)
(0.795918367346939,0.00311505386061131)
(0.816326530612245,0.00243313395988485)
(0.836734693877551,0.00185486319988541)
(0.857142857142857,0.00137092083145563)
(0.877551020408163,0.000972824636318401)
(0.897959183673469,0.000652855425612568)
(0.918367346938776,0.000403988348403636)
(0.938775510204082,0.000219830393692436)
(0.959183673469388,9.45635256950274e-05)
(0.979591836734694,2.28929431850364e-05)
(1,5.55111512312578e-17)

};
\path [draw=black, fill opacity=0] (axis cs:13,3.64161943162664)--(axis cs:13,3.64161943162664);

\path [draw=black, fill opacity=0] (axis cs:1,13)--(axis cs:1,13);

\path [draw=black, fill opacity=0] (axis cs:13,5.55111512312578e-17)--(axis cs:13,5.55111512312578e-17);

\path [draw=black, fill opacity=0] (axis cs:0,13)--(axis cs:0,13);

\end{axis}

\end{tikzpicture}